\documentclass[abbrv,aps,prl,twocolumn,showpacs,preprintnumbers,amsmath,amssymb,
superscriptaddress,nofootinbib]{revtex4}

\usepackage{bbm}
\usepackage[english]{babel}
\usepackage[latin1]{inputenc}
\usepackage{amssymb,amsmath}
\usepackage{ushort}

\usepackage{graphicx} 
\usepackage{dcolumn}
\usepackage{bm}

\begin{document}
\title {A Microscopic View on the Mott transition in Chromium-doped V$_2$O$_3$
}

\author{S.~Lupi}
\affiliation {CNR-IOM and Dipartimento di Fisica, Universit\`a di Roma "La Sapienza",
Piazzale A. Moro 2, I-00185 Roma, Italy}

\author{L.~Baldassarre}
\affiliation {Sincrotrone Trieste S.C.p.A., Area Science Park, I-34012 Basovizza, Trieste, Italy}

\author{B.~Mansart} 
\affiliation {Laboratoire de Physique des Solides, CNRS-UMR 8502, Universit\'{e} Paris-Sud, F-91405 Orsay, France}

\author{A.~Perucchi}
\affiliation {Sincrotrone Trieste S.C.p.A., Area Science Park, I-34012 Basovizza, Trieste, Italy}

\author{A.~Barinov}
\affiliation {Sincrotrone Trieste S.C.p.A., Area Science Park, I-34012 Basovizza, Trieste, Italy}

\author{P.~Dudin} 
\affiliation {Sincrotrone Trieste S.C.p.A., Area Science Park, I-34012 Basovizza, Trieste, Italy}

\author{E.~Papalazarou} 
\affiliation {Laboratoire de Physique des Solides, CNRS-UMR 8502, Universit\'{e} Paris-Sud, F-91405 Orsay, France}

\author{F.~Rodolakis} 
\affiliation {Laboratoire de Physique des Solides, CNRS-UMR 8502, Universit\'{e} Paris-Sud, F-91405 Orsay, France}
\affiliation {Synchrotron SOLEIL, L'Orme des Merisiers, Saint-Aubin, BP~48, 91192 Gif-sur-Yvette Cedex, France}

\author{J.-P.~Rueff} 
\affiliation {Synchrotron SOLEIL, L'Orme des Merisiers, Saint-Aubin, BP~48, 91192 Gif-sur-Yvette Cedex, France}

\author{J.-P.~Iti\'{e}} 
\affiliation {Synchrotron SOLEIL, L'Orme des Merisiers, Saint-Aubin, BP~48, 91192 Gif-sur-Yvette Cedex, France}

\author{S.~Ravy} 
\affiliation {Synchrotron SOLEIL, L'Orme des Merisiers, Saint-Aubin, BP~48, 91192 Gif-sur-Yvette Cedex, France}

\author{D.~Nicoletti} 
\affiliation {Dipartimento di Fisica, Universit\`a di Roma "La Sapienza",
Piazzale A. Moro 2, I-00185 Roma, Italy}

\author{P.~Postorino}
\affiliation {CNR-IOM and Dipartimento di Fisica, Universit\`a di Roma "La Sapienza",
Piazzale A. Moro 2, I-00185 Roma, Italy}

\author{P.~Hansmann} 
\affiliation {Institute of Solid State Physics, Vienna University of Technology, 1040 Vienna, Austria}

\author {N.~Parragh}
\affiliation {Institute of Solid State Physics, Vienna University of Technology, 1040 Vienna, Austria}

\author{A.~Toschi} 
\affiliation {Institute of Solid State Physics, Vienna University of Technology, 1040 Vienna, Austria}

\author{T.~Saha-Dasgupta}  
\affiliation{S.N.Bose Center for Basic Sciences, Salt Lake, Kolkata, India}

\author{O.~K.~Andersen} 
\affiliation{Max-Planck-Institut f\"ur Festk\"orperforschung, Heisenbergstrasse 1, D-70569 Stuttgart, Germany}

\author{G.~Sangiovanni} 
\affiliation {Institute of Solid State Physics, Vienna University of Technology, 1040 Vienna, Austria}

\author{K.~Held} 
\affiliation {Institute of Solid State Physics, Vienna University of Technology, 1040 Vienna, Austria}

\author{M.~Marsi}
\affiliation {Laboratoire de Physique des Solides, CNRS-UMR 8502, Universit\'{e} Paris-Sud, F-91405 Orsay, France}

\date{\today}

\maketitle

\textbf{
V$_2$O$_3$ is the prototype system for the Mott transition, one of the most fundamental phenomena of electronic correlation. Temperature, doping or pressure induce a metal to insulator transition (MIT) between a paramagnetic metal (PM) and a paramagnetic insulator (PI). This or related MITs have a high technological potential, among others for intelligent windows and field effect transistors. However the spatial scale on which such transitions develop is not known in spite of their importance for research and applications. Here we unveil for the first time the MIT in Cr-doped  V$_2$O$_3$ with submicron lateral resolution: with decreasing temperature, microscopic domains become metallic and coexist with an insulating background. This explains why the associated PM phase is actually a poor metal. The phase separation can be associated with a thermodynamic instability near the transition. This instability is reduced by pressure which drives a genuine Mott transition to an eventually homogeneous metallic state.
}
   
Hitherto, the phase diagram of V$_2$O$_3$, see Fig.\ref{Fig1}(a), was mainly established by means 
of transport measurements, i.e. detecting the electronic response on a $\it{macroscopic}$ spatial scale \cite{McWhan}. According to this phase diagram, V$_2$O$_3$ exhibits a first-order MIT from a low-$T$ antiferromagnetic insulating state (AFI) to a paramagnetic metallic state (PM) when $T$ exceeds $T_N \sim 160$ K.  A different, paramagnetic insulating state (PI) is reached at $T>T_N$ if V is partially substituted by Cr. Within a narrow range of Cr content  (0.005$<x<$0.018) a PI to PM transition is obtained upon reducing the temperature across a $T_{\rm MIT}>T_N$ with a slight discontinuity of the lattice parameters within the same corundum phase.
In particular  (V$_{0.989}$Cr$_{0.011}$)$_2$O$_3$ has attracted great interest since its discovery in 1969 \cite{McWhan}, because it provides the opportunity to span all phases, through two MITs. One (AFI $\leftrightarrow$ PM) shows both a change of crystal symmetry and of magnetic state, the other (PM $\leftrightarrow$ PI) is iso-structural and a prototype of the correlation-driven Mott transition.

It was soon evident that this simple scheme hides a more complicated scenario.  
In their conducting phase weakly Cr-doped samples show a poor-metal behavior (resistivity $\sim$ 10$^{-2}$ $\Omega$  cm) \cite{McWhan, kuwamoto}. Moreover the structure of (V$_{0.989}$Cr$_{0.011}$)$_2$O$_3$  is described as coexistence of an $\alpha$-phase, with lattice parameters close to those of metallic V$_2$O$_3$ at 400 K, and a $\beta$-phase with basically the same
structure as at higher  Cr content \cite{remeika,robinson}. In addition, EXAFS measurements \cite{frenkel} suggest the essential role played by the local lattice strain which occurs in the Cr-doped compounds: Cr contracts the Cr-V bond, inducing a concomitant elongation of the V-V distances \cite{frenkel}.
As shown by theoretical calculation based on the Local-Density Approximation combined with Dynamical Mean Field Theory (LDA+DMFT) \cite{held}, 
a longer V-V bond is associated with the PI insulating phase \cite{remeika}.
Besides these open issues concerning the T-induced MIT, further
questions are raised by studies under external pressure: in (V$_{0.989}$Cr$_{0.011}$)$_2$O$_3$, for instance, a conducting phase is obtained at room temperature if $P$ $>$ 3 kbar \cite{limelette}, and early investigations stated that applying pressure is equivalent to decreasing the Cr content \cite{jayaraman}. Recently it was observed that different electronic structures for the metallic phase can be obtained when the PI-PM transition is induced by doping temperature or pressure \cite{Rodolakis2}, challenging the simple scenario of the 
McWhan phase diagram \cite{McWhan}.   

In this paper we explore the phase diagram of Cr-doped V$_2$O$_3$ from $\it{macroscopic}$ to $\it{microscopic}$ scales, by combining three spectroscopies with different lateral resolution,  i.e., Infrared (IR), Scanning Photoemission Microscopy (SPEM), and X-ray Diffraction (XRD), with LDA+DMFT calculations. We find unambiguous evidence of an electronic phase separation in the T-induced PM phase. The poor-metallic behaviour is therefore explained in terms of a coexistence of metallic and insulating domains stabilized by structural defects. Both infrared and XRD measurements demonstrate that external pressure can drive the MIT much closer to a genuine Mott transition, leading eventually to a homogeneous metallic phase, as expected insofar in (V$_{0.989}$Cr$_{0.011}$)$_2$O$_3$. However, to recover the lattice parameters and the electronic structure of undoped V$_2$O$_3$ a much higher pressure is needed than expected from macroscopic transport measurements \cite{McWhan, limelette}. These findings shine new light on the properties of vanadium sesquioxide. In particular the phase separation occurring along some pathways across the PI-PM transition translates into intrinsically different properties of the corresponding PM phase.

\section{Results}
In the PM state at 220 K, the optical conductivity $\sigma_1(\omega)$ of (V$_{0.989}$Cr$_{0.011}$)$_2$O$_3$, reported in Fig.\ref{Fig1}(b) shows a pseudo-gap in the far infrared but absolutely no Drudelike contribution.
One may also notice that the extrapolation to $\omega$=0 is in excellent agreement with the $dc$ conductivity. The absence of a Drude term is in striking contrast with what is seen in V$_2$O$_3$ at the same temperature (Fig.\ref{Fig1}(c)), and implies that within the PM region of the phase diagram the metallic properties can change dramatically.  On the contrary, in the PI phase  the spectrum of (V$_{0.989}$Cr$_{0.011}$)$_2$O$_3$ is akin to that of (V$_{0.972}$Cr$_{0.028}$)$_2$O$_3$: both spectra are characterized by a small insulating gap associated with an absorption band centered around 3000 cm$^{-1}$. The T-induced PI-PM transition thus corresponds only to a weak red-shift of the spectral weight (Fig.\ref{Fig1}(b)). The optical data also show a strong thermal hysteresis (Fig.\ref{Fig1}(b)): upon cooling the MIT occurs below $\sim$ 260 K; upon heating thereafter the sample, the PM phase remains stable up to 275 K, and also at 300 K the PI phase is not fully recovered. 

The absence of a metallic contribution in the optical spectra of the x=0.011 sample calls for further investigations, especially since recent photoemission measurements on crystals from the same batch detected a clear metallic quasiparticle peak \cite{skMo,skMoPRB,Rodolakis,evangelos}, when performed with sufficient bulk sensitivity \cite{borghi}. To better understand this apparently contradictory results, we performed SPEM with sub-micrometric spatial resolution \cite{Dudin}, revealing an inhomogeneous electronic distribution in the PM phase.
The maps of the photoelectron yield at $E_F$ \cite{Sarma, Gunther} are plotted in Fig. \ref{Fig2} together with representative photoemission spectra.   
It should be pointed out that our experimental geometry \cite{Dudin} and  photon energy (27 eV) did not allow the collection of  sufficient bulk sensitive photoelectron spectra \cite{skMo,skMoPRB,Rodolakis,evangelos}. Thus the spectra corresponding to the metallic phase do not show a pronounced quasiparticle peak. Nevertheless, the difference between insulating and metallic phases is unambiguous. Consequently, we interpret the photoemission microscopy images as representative of the topmost part of the metallic and insulating domains, whose thickness is expected to be comparable to the IR probing depth. This penetration depth is sufficient to provide the excellent agreement between optical and the dc conductivity obtained with transport measurements.
With these maps we resolve metallic and insulating domains which appear as $T\!<\!T_{\rm MIT}$ and persist in the PM phase. A uniform insulating behavior is found instead in the PI phase (Fig.\ref {Fig2}(c)), showing that the PM state only is strongly phase separated. 
Furthermore, when the sample is cooled down again (Fig. \ref{Fig2}(d)), the metallic and insulating domains are still found in the same position 
and with a similar shape, clearly showing that they form around 
specific points in the sample. It is natural to correlate these nucleation centers to structural inhomogeneites in the material. Such a tendency towards phase separation is related to the correlation energy which fosters a thermodynamic instability in the vicinity of the Mott transition \cite{TDinst1,TDinst2}.This is anologous to  the liquid-vapour transition, and we can hence visualize the PI-PM transition as the formation of metallic clouds within an insulating sky.

Although a T-induced structural phase separation was  previously proposed
on the basis of diffraction \cite{McWhan,robinson} and EXAFS \cite{frenkel} experiments, it was never associated before with an inhomogeneous electronic state. Our data maps for the first time the MIT on the microscale, showing metallic and insulating domains which can be naturally linked to the $\alpha$ and $\beta$ lattice structures, respectively.

On the basis of these experimental findings in photoemission microscopy we can also understand the poor-metallic behavior in the PM phase, detected in the optical spectra. To further elaborate on this, we employ the Effective Medium Approximation (EMA)\cite{tanner} which is widely recognized to be meaningful in the presence of microscopic inhomogeneities smaller than the wavelength of the electromagnetic radiation. EMA describes the (complex) dielectric function $\tilde\varepsilon_E$ of an inhomogeneous system \cite{Bruggeman, Basov} in terms of the optical response of the two components (details in the supplementary material), weighted by the respective volume fraction $f$ and depolarization factor $q$:

\begin{displaymath}
f \frac{\tilde{\varepsilon}_{\alpha}-\tilde{\varepsilon}_{E}}{\tilde{\varepsilon}_{\alpha}+\frac{1-q}{q}\tilde{\varepsilon}_{E}} + (1-f) \frac{\tilde{\varepsilon}_{\beta}-\tilde{\varepsilon}_{E}}{\tilde{\varepsilon}_{\beta}+\frac{1-q}{q}\tilde{\varepsilon}_{E}} = 0
\label{equation}
\end{displaymath}
Here, $\tilde{\epsilon}_{\alpha}$ and $\tilde{\epsilon}_{\beta}$ are the complex optical dielectric functions of the metallic and the insulating phase, respectively.

We reproduce the experimental conductivity in the PM phase of (V$_{0.989}$Cr$_{0.011}$)$_2$O$_3$ at any $T$ by using as input parameters the $\alpha$ and $\beta$ conductivities as measured in V$_2$O$_3$ \cite{Baldassarre} and (V$_{0.972}$Cr$_{0.028}$)$_2$O$_3$, respectively. 
In particular previous measurements showed that the optical properties of V$_2$O$_3$ were strongly dependent on lattice modifications \cite{Baldassarre}. As the lattice parameters of $\alpha$ (V$_{0.989}$Cr$_{0.011}$)$_2$O$_3$ phase correspond to those of V$_2$O$_3$ at 400 K \cite{McWhan,robinson}, we choose the corresponding 400 K optical conductivity for the EMA reconstruction. Instead, the optical conductivity of (V$_{0.972}$Cr$_{0.028}$)$_2$O$_3$, representative of the $\beta$ phase, being nearly T-independent was choosen at 220 K. 
Extracting $f$= 0.45$\pm$0.05 from the map in Fig.\ref{Fig2}(a) (see Supplementary Figure S1 and discussion in the supplementary material), and leaving $q$ as the only free-fitting parameter a good agreement between fit (empty squares in Fig.\ref{Fig1}(d)) and experimental data (solid line) is found for $q$= 0.30$\pm$0.05 at 220 K ($q$= 0.40$\pm$0.05 at 300 K), similar to the value found in VO$_2$ across its first order MIT  \cite{Basov}. Since $f\!>\!q$ the MIT transition is percolative in nature, thus explaining the low dc values measured in samples coming from the same batch. Infrared measurements associated with SPEM data therefore clarify the poor-metallic behavior observed on a $\it{macroscopic}$ spatial scale.

To further support the paramount role played by phase separation for
the low-energy excitations of (V$_{0.989}$Cr$_{0.011}$)$_2$O$_3$, we compare our infrared results with state-of-the art LDA+DMFT  calculations \cite{silke, paper2} (Supplementary Figure S2 and discussion in the supplementary material).
The LDA+DMFT optical conductivity displays a metallic peak in the 
PM $\alpha$ phase (yellow line) and a gap 
followed by the $t_{2g}\rightarrow t_{2g}$ excitations
in the insulating $\beta$ PI phase (gray line) (Fig.\ref{Fig1}(e)). 
To calculate the LDA+DMFT optical conductivity for the PM phase at x=0.011 we
follow what our experimental analysis suggests, namely to take a mixture of
the pure $\alpha$- and $\beta$-phases. An important point is that, while EMA
in the experimental analysis is (necessarily) performed by assuming the
metallic phase to be represented by x=0 and the insulating one by x=0.028, our
LDA+DMFT approach allows us to use the "actual" $\alpha$- and $\beta$-phases at x=0.011 in the EMA formula. Using the same
values for the parameters $f$ and $q$ for theory and experiment, we find a
LDA+DMFT optical conductivity at x=0.011 (open symbols in Fig.\ref{Fig1}(e))
which strongly resembles the experimental curve: in particular, its behavior
is markedly insulating-like, as in the experiment.
Note that tendencies towards phase separation close to a MIT have been discovered in DMFT before,
on the basis of a negative compressibility 
\cite{TDinst1} and a negative curvature of the energy vs. volume curve \cite{TDinst2}. These tendencies can be further enhanced by the coupling of electronic and lattice degrees of freedom.

Let us now turn to the P-induced MIT: at the minimum $P$ (1 kbar), $\sigma_1(\omega)$ shows  an overall insulating behavior resembling
that at ambient pressure. As $P$ is increased, $\sigma_1(\omega)$ changes abruptly. Indeed, a sudden transfer of spectral weight from above $\sim$3000 cm$^{-1}$ to lower frequency builds up a metallic conductivity which well extrapolates to the dc values (solid diamonds in Fig.\ref{Fig3}), measured on samples of the same batch. The difference from the  
poor-metallic behavior found in the T-induced PM phase at 220 K (and ambient-P) is striking.
Nevertheless, even at 15 kbar, the optical conductivity of V$_2$O$_3$ is not
yet recovered, challenging the proposed scaling relation between Cr-doping and
pressure ($x\sim$ 0.011 $\approx$ P$\sim$ 4 kbar) obtained by dc resistivity measurements \cite{jayaraman,limelette}.
The metallicity induced by pressure can be associated with a double process: a weaker phase separation and a modification of the $\alpha$-domain lattice parameters, which may become more similar to those of pure V$_2$O$_3$ at room temperature.

In order to clarify these issues, high pressure XRD measurements have been performed in the same pressure range and the results for the (300) Bragg line (confirmed also by the similar behavior on other Bragg peaks) are shown in Fig.\ref{Fig4}(a). Two Bragg peaks appear within a range of
$\sim\!1\,$ kbar across the MIT, demonstrating that the PI phase coexists with the PM one only in this limited pressure range. A further increase of $P$ strongly reduces the phase separation (see Fig.\ref{Fig4}(b)) and above 10 kbar the Bragg signal is made almost entirely by the PM component.

These data indicate that a more homogenous electronic state develops in the P-induced PM phase. Indeed, an EMA calculation reproduces the P-dependent optical conductivity (see inset of Fig.\ref{Fig3}), through the same input curves as before. For the same value of $q$=0.4 obtained at ambient P and room T, a metallic volume fraction $f$ =0.85 at 6 kbar and 0.95 at 15 kbar can be found. These values (solid symbols in Fig.\ref{Fig4}(b)) are in good agreement with values obtained from XRD measurements, but much larger than the metallic fraction $f$ = 0.42 found at 220 K in the T-induced PM phase through SPEM. Eventually, a metallic volume fraction $f\sim$1 is found at pressures where the $c/a$ value in Fig.\ref{Fig4}(c) nearly attains that of pure V$_2$O$_3$ ($c/a\sim$2.83), thus suggesting the key role played by the lattice in the physical properties of V$_2$O$_3$ \cite{Baldassarre, Finger}.

\section{Discussion}
We explored the phase diagram of (V$_{0.989}$Cr$_{0.011}$)$_2$O$_3$ with a combination of different experimental methods and conjoined LDA+DMFT calculations, with particular attention to the PI-PM transition which can be obtained by changing doping, temperature or pressure. Our results  shine new light on the different metallicity of the PM phase obtained by lowering $T$ with respect to the one obtained by applying an external P. We showed that a key role to explain this difference is played by phase separation: in particular, the coexistence of metallic and insulating domains was unambiguously demonstrated in the T-induced PM phase. After a thermal cycle these domains reappear in the same positions and with a similar shape: this memory effect may be related to nucleation centers, with electronic correlations further fostering phase separation. The PM phase reached under pressure is instead much more homogeneous and, by increasing P, it soon covers a raising fraction of the volume. However, it is intrinsically different from pure V$_2$O$_3$ if one compares points which are nominally equivalent in the McWhan phase diagram, as demonstrated by the different lattice parameters. 
This suggests that, during the phase transition induced by P, the strain between the different lattice parameters between PM and PI seems to be accomodated in a more homogeneous way by reducing their difference. In the T-induced transition, instead, this difference is unchanged and the strain is accomodated inhomogeneously inducing domains with different lattice and electronic properties. 

Our infrared results, obtained by changing doping, temperature and pressure, were captured correctly by employing the effective medium approximation. Thus, optical conductivity bridges between local probes like photoemission and XRD, which can directly detect the microscopic details of the phase separation, and transport measurements, which are instead sensitive to the occurance of a long scale metallic property. Thanks to this combination of different methods with intrinsically different spatial scale sensitivity, we were able to clarify the properties of the phase diagram of vanadium sesquioxide from macroscopic to microscopic spatial scales, showing that the phase separation occuring along the different pathways across the PI-PM transition translates into different properties of the PM phase.  

In conclusion, our  investigation shows that the phase diagram of V$_2$O$_3$, 
celebrated for its essential simplicity, reveals on a microscopic 
scale a much richer phenomenology which has to be taken into 
account in the physical description of this model system.

\section{Methods}
\subsection{Materials} 
All the measurements on (V$_{0.989}$Cr$_{0.011}$)$_2$O$_3$, employing different experimental techniques as detailed in this paper, were performed on single crystals from the same batch grown at Purdue University. The specimens were thoroughly characterized prior to the experiments by means of transport measurements (Patricia Metcalf, private communication), and precisely oriented with X-ray diffraction.
\subsection{Infrared Measurements}
The temperature-dependent reflectivity (R) measurements have been performed at near-normal incidence with the electrical field polarized within the ab plane and by using as a reference a metallic film (gold or aluminium depending on the spectral range) evaporated $in$ $situ$ over the sample surface. A standard low-frequency extrapolation was performed by using a Drude-Lorentz fitting of the measured R by taking into account the dc values of the resistivity measured on samples of the same batch, while the high-frequency tail from Ref.\cite{Baldassarre} was merged to the data, in order to perform reliable Kramers-Kronig transformations. A diamond anvil cell (DAC) equipped with type IIa diamonds was used to perform pressure-dependent measurements. A hole of about 300 $\mu$m diameter was drilled in a 50 $\mu$m thick Al gasket, allowing us to span over the desired pressure range (0-15 kbar) in a reproducible  and cyclic way. A small piece was cut from the sample measured as a function of T and placed in the sample chamber together with a KBr pellet (used as hydrostatic medium) and the electric field of the radiation was polarized within the ab plane. We took great care of creating a clean sample diamond interface. The pressure was measured \emph{in situ} with the standard ruby fluorescence technique \cite{insitu}. In order to reduce the experimental error of such data, a ruby micro-sphere was placed inside the sample chamber while a second one was placed on the external face of the diamond. Measurements were performed on both ruby balls several times for each pressure, taking therefore in account any slight change in the temperature or in the spectrometer calibration. The incident and reflected light were focused and collected with a cassegrain-based optical microscope equipped with a MCT detector and a bolometer and coupled to a Michelson interferometer. This allowed us to explore the 350-12000 cm$^{-1}$ spectral range exploiting the high brilliance of Synchrotron Radiation at SISSI beamline at ELETTRA storage ring \cite{sissi}. The measurement  procedure was the same as described in Refs. \cite{v3o5,sacchetti}. The conductivity, $\tilde{\sigma}(\omega)$, has been calculated through Kramers-Kronig transformations, taking into account the diamond-sample interface \cite{plaskett,kuntscher,hemely}. Data are not shown in the 1600-2700 cm$^{-1}$ due to the strong phonon absorption of diamonds.
\subsection{Photoemission Microscopy}
The photoemission microscopy results were obtained on the Spectromicroscopy beamline
\cite{Dudin} at the Elettra. Photons at 27 eV were focused through a Schwarzschild objective,  to obtain a submicron size spot.  The measurements were performed in ultra-high vacuum (P$<2$$\times10^{-10}$ mbar) on in situ prepared surfaces of (V$_{0.989}$Cr$_{0.011}$)$_2$O$_3$. Photoelectrons were  collected at an emission angle of 65 degrees from the normal, which makes the photoemission spectra and images surface sensitive.
A standard photoemission microscopy procedure was employed to remove topographic features \cite{marsi_jesrp} from the images presented in Fig.\ref{Fig2}, and to obtain an unambiguous contrast between metallic and insulating regions \cite{Sarma}: the spectral intensity at the Fermi level E$_F$ - high in the PM and absent in the PI phase \cite {skMoPRB, Rodolakis} - was divided by the intensity recorded at the V3d-band. Since the latter remains essentially constant between the PI and PM phases, two-dimentional maps of this ratio give a genuine representation of the lateral variation of metallic and insulating domains. 
\subsection{X-ray diffraction}
The X-ray diffraction measurements were performed using 17 keV photons dispersed by the CRISTAL beamline at the SOLEIL synchrotron radiation  
source \cite{Cristal}. Pressure was applied to a (V$_{0.989}$Cr$_{0.011}$)$_2$O$_3$ powder by means of a membrane driven diamond anvil cell, and its value in the cell was carefully measured during the measurement sequence using the standard ruby fluorescence method. The transmission of the diamond cell at the photon energy employed during these experiments was sufficient to allow a high resolution detection of the main Bragg reflections like the (300) and (214) presented here.

\section{Acknowledgements}
We acknowledge financial support from the European Community's Seventh 
Framework Programme
(FP7/2007-2013) under grant agreement n¼ 226716,  from the RTRA 
"Triangle de la Physique", and the Austrian Science Fund (FWF) through WK004.
We thank P. Metcalf for providing (V$_{1-x}$Cr$_{x}$)$_2$O$_3$  single crystals of excellent 
quality, and P. Calvani,
P. Wzietek and D. J\'{e}rome for stimulating discussions.

\section{Authors Contributions}
L.B. and B. M. carried out the infrared and the photoemission microscopy
experiments and data analysis, with the help of A.P., 
A.B., P.D. D.N. and E.P.
F.R., J-P.R., J-P.I. and S.R. performed the XRD measurements, P.H., 
N.P., A.T., T.S-D.,
O.K.A., G.S. and K.H. the LDA+DMFT calculations. S.L. and M.M.
were responsible for the planning and the management of the project with inputs from all the co-authors, especially from L.B., A.P., 
B.M. and K.H. All authors extensively 
discussed the results and the
manuscript that was written by S.L., L.B. and M.M. 

\section{Additional Information}
The authors declare no competing financial interests. Supplementary material accompanies this paper on http://www.nature.com/naturecommunications.
Correspondence and requests for materials should be adressed to S.L. (stefano.lupi@roma1.infn.it) and M.M. (marino.marsi@u-psud.fr).

\begin{widetext}

\begin{figure}[h]   
\begin{center}    
\includegraphics[width=18cm]{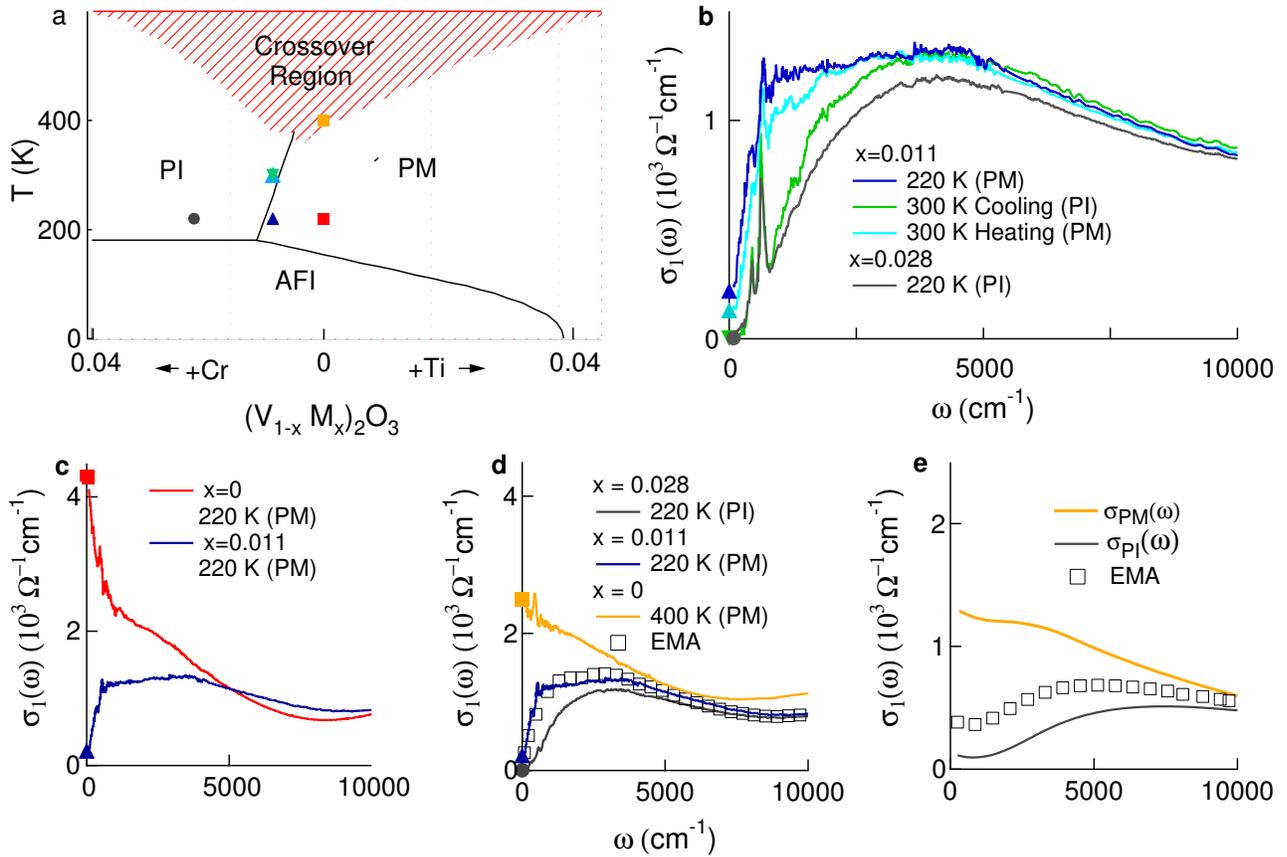}
\caption{ {\bf Temperature dependent infrared measurements.}
(a) Schematic phase diagram as defined in Ref.\ \cite{McWhan};
markers indicate the measurements displayed
in (b)-(e) in the same color coding.  b) $\sigma_1(\omega)$ of (V$_{1-x}$Cr$_{x}$)$_2$O$_3$ for x=0.011 and 0.028 at selected $T$. $\sigma_1(\omega)$ for x=0.011 displays different behaviors in the PI and PM regions. The two curves measured at 300 K highlight the presence of large hysteresis: the curve at 300 K reached from the PI side (upon cooling) shows the expected insulating behavior that is not fully recovered when 300 K is reached from the PM phase (upon heating). In the PI phase only a small difference is found in $\sigma_1(\omega)$ of x=0.011 and 0.028. Solid symbols in panels $b$ to $d$ represent dc values as measured in Ref.\cite{limelette} on crystals of the same batch. (c) $\sigma_1(\omega)$ in the PM phase highlighting the difference between
V$_2$O$_3$ and (V$_{0.989}$Cr$_{0.011}$)$_2$O$_3$
showing a clear Drude peak and a poor-metal behavior, respectively. (d) Experimental and (e) theoretical reconstruction of 
$\sigma_1(\omega)$ in the T-induced PM phase, succesfully obtained
by modelling, with the same $q$ and $f$ parameters (see text), the coexistence of metallic and insulating domains
through the effective medium approximation (EMA).}

\label{Fig1} 
\end{center}
\end{figure}

\begin{figure}[h]
\begin{center}
\includegraphics[width=16cm]{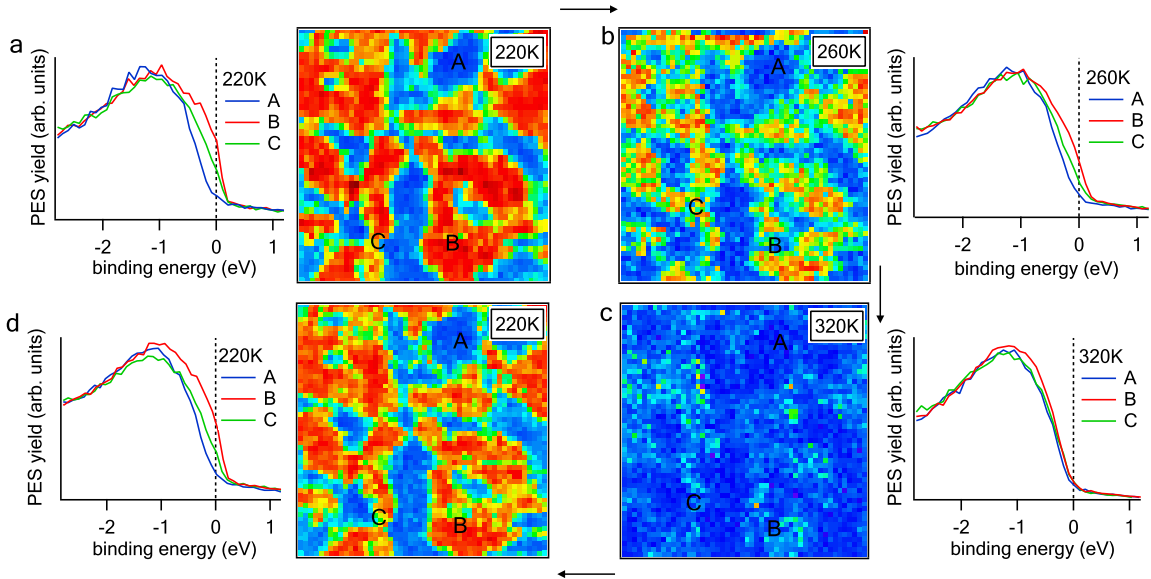}\\
\caption { {\bf Temperature dependent photoemission microscopy measurements.}
Scanning photoemission microscopy images and spectra on (V$_{0.989}$Cr$_{0.011}$)$_2$O$_3$, collected at 27 eV photon energy and at different temperatures on a 50 $\mu$m by 50 $\mu$m sample area. The images are obtained with standard procedures to remove background and topography effects \cite{Sarma}. The pictorial contrast between metallic and insulating zones is obtained from the photoemission intensity at the Fermi level \cite{Gunther}, normalized by the intensity on the V3d band (binding energy -1.2 eV). 
The lateral variation of this normalized intensity is represented in the color scale (the intensity scale is the same for all four maps) from its  minimum value, in violet, to its maximum, in red, and is therefore a direct visualization in real space of the metallicity of the system. 
Inhomogeneous properties are found within the PM phase at T=220 K (a) and 260 K (b), where metallic (in red) and insulating (in blue) domains coexist. To recover the PI phase, the sample was heated to 320 K, to make sure that the hysteresis effects still present at 300 K (see Fig. 1) are overcome: at this temperature (c) a homogenous insulating state is obtained. After a whole thermal cycle (d) the metallic regions can be found in the same position and shape as in (a). The photoemission spectra (outer panels) from selected representative areas (A,B,C) corroborate this interpretation.}

\label{Fig2}
\end{center}
\end{figure}

\begin{figure}[h]   
\begin{center}    
\includegraphics[width=8.5cm]{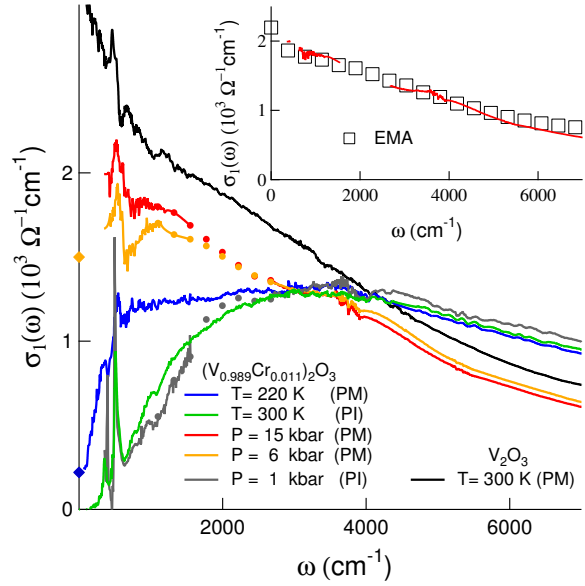}  
\caption{ {\bf Pressure dependent infrared measurements.}
Optical conductivity of (V$_{0.989}$Cr$_{0.011}$)$_2$O$_3$ under pressure (data between 1600 and 2700 cm$^{-1}$, due to 
the strong diamond phonon absorption, are substituted by markers as a guide for the eye). Under increasing P, the optical conductivity is depleted at energies above 3000 cm$^{-1}$ and increases below this energy.  The curves at 1 kbar, 6 kbar and 15 kbar are compared to the PM state of V$_2$O$_3$ at 300 K and to the poor metallic phase of (V$_{0.989}$Cr$_{0.011}$)$_2$O$_3$ at 220 K. In the latter phase no metallic-like increase 
is built up for  $\omega\rightarrow 0$, whereas as a function of pressure some coherent transport takes place, as supported also from the resistivity data (solid diamonds in Fig.\ref{Fig3}) that allows to extrapolate smoothly  the curves to $\omega\rightarrow0$. In the inset is shown the experimental reconstruction of $\sigma_1(\omega)$ in the P-induced PM phase obtained by EMA.
} 
\label{Fig3}
\end{center}
\end{figure}

\begin{figure}[h]   
\begin{center}    
\includegraphics[width=8.5cm]{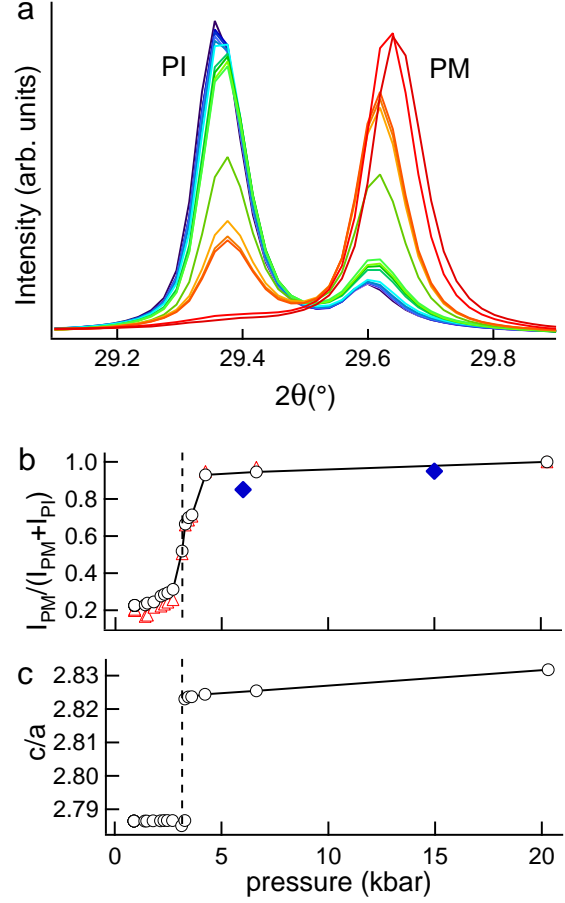}  
\caption{  {\bf Pressure dependent Bragg X-ray diffraction measurements.}
(a)  Bragg X-ray diffraction (300) line of (V$_{0.989}$Cr$_{0.011}$)$_2$O$_3$ under pressure at room temperature. The peak at the left corresponds to the lattice parameters of the PI 
phase, while the one at the right corresponds to PM. The spectra from violet to red
were taken at pressures from 0 to 20 kbar, corresponding to the data points in Fig.\ref{Fig4}(b) and (c). (b)  PM  intensity over the total one (PM+PI)  as a function of $P$ for the (300) line (circles) and for the (214) line (triangles). Solid symbols represent the same quantity calculated from the $f$ EMA parameter (see above). (c) Ratio of the lattice parameter c/a vs.\ $P$   extracted from  Fig.\ref{Fig4}(a).
} 
\label{Fig4}
\end{center}
\end{figure}

\end{widetext}

\clearpage




\noindent{Supplementary material}\\

\Large{A Microscopic View on the Mott transition in Chromium-doped V$_2$O$_3$}\\

\noindent\small{S.~Lupi, L.~Baldassarre, B.~Mansart, A.~Perucchi, A.~Barinov, P.~Dudin, E.~Papalazarou, F.~Rodolakis, J.-P.~Rueff, J.-P.~Iti\'{e}, S.~Ravy, D.~Nicoletti, P.~Postorino, P.~Hansmann, N.~Parragh, A.~Toschi, T.~Saha-Dasgupta, O.~K.~Andersen,  
G.~Sangiovanni, K.~Held and M.~Marsi}\\


\noindent\Large{Effective medium approximation}\\

\noindent\small{Within the Bruggeman EMA theory (Bruggeman, D.A.G., {\it Ann. Phys. Lpz.}, {\bf 24}, 636 (1935)), which has been successfully used to describe the optical properties of phase-separated correlated systems (Ref.22, main paper), the complex optical dielectric function ($\tilde{\varepsilon}_{E}$) is described by the following equation:
\begin{displaymath}
f \frac{\tilde{\varepsilon}_{\alpha}-\tilde{\varepsilon}_{E}}{\tilde{\varepsilon}_{\alpha}+\frac{1-q}{q}\tilde{\varepsilon}_{E}} + (1-f) \frac{\tilde{\varepsilon}_{\beta}-\tilde{\varepsilon}_{E}}{\tilde{\varepsilon}_{\beta}+\frac{1-q}{q}\tilde{\varepsilon}_{E}} = 0
\label{equation}
\end{displaymath}
where $\tilde{\epsilon}_{\alpha}$ ($f$) and $\tilde{\epsilon}_{\beta}$ (${1-f}$) are the complex optical dielectric functions (the filling factors) of the metallic and insulating phase respectively and $q$ is the depolarization factor that depends on the shape of the metallic domains dispersed in the insulating matrix.  

We reproduce the experimental conductivity in the PM phase of (V$_{0.989}$Cr$_{0.011}$)$_2$O$_3$ at any $T$ by using as input parameters the $\alpha$ and $\beta$ conductivities as measured in V$_2$O$_3$ (Ref. 23, main paper) and (V$_{0.972}$Cr$_{0.028}$)$_2$O$_3$ (Fig. 1b in the paper), respectively.

The filling factor was determined from the photoemission images of the spectral weight at E$_{F}$ (Fig.2 in the paper). 
In particular the bimodal histogram of the intensity distribution extracted from the PM map (Supplementary Fig.S1) shows unambiguously that at 220 K two main phases can be identified: one insulating, corresponding to the violet-blue colors in the photoemission map, and one metallic, corresponding to orange-red colors. The broad minimum between these two components corresponds to the green color in the microscopic maps, which we interpret as regions where metallic and insulating regions are too small to be resolved with our instrumental resolution.These considerations lead us to conclude that the assumption of a two-component description for the PM phase is perfectly legitimate. Furthermore, the PM histogram can be fitted by a sum of two Gaussian curves (dashed line), whose relative intensities provide the insulating and metallic filling factors. Specifically, the fit provides for the metallic part f=0.45$\pm$0.05.
Remarkably, at 320 K only (not shown) one Gaussian (corresponding to the insulating phase) is sufficient to fit the intensity distribution histogram, as expected from the corresponding photoemission microscopy map (Fig. 2c in the paper), which shows an almost uniform blue-violet color.  

As the EMA analysis is strongly constraint by the pseudogap behavior of the optical conductivity below 1000 cm$^{-1}$ (see Fig.1d in the paper) as well as by the finite (independently measured) dc value ($\sigma$(dc) $\sim$ 250 $\Omega^{-1}$ cm$^{-1}$), by changing $f$ in the EMA analysis accordly to the previous estimate, the reconstruction is still possible with a similar quality for different values of $q$ (as far $f\!>\!q$ in order to have a finite dc conductivity). From f=0.45$\pm$0.05 one can estimate q=0.30$\pm$0.05.}\\

\noindent\Large{LDA+DMFT Optical conductivity}\\

\noindent\small{The formula for the optical conductivity reads\\
\begin{widetext}
\begin{equation}\label{finalopt}
\Re\sigma^{\alpha\beta}(\Omega)=\frac{2\pi
  e^2\hbar}{V}\sum_{\mathbf{k}}\int
d\omega'\frac{f(\omega')-f(\omega'+\Omega)}{\Omega}\;\text{Tr}\left[\underline{\underline{v}}_\alpha(\mathbf{k})\underline{\underline{A}}(\mathbf{k},\omega')\underline{\underline{v}}_\beta(\mathbf{k})\underline{\underline{A}}(\mathbf{k},\omega'+\Omega)\right]
\end{equation}\\
\end{widetext}
where $V$ is the volume of the unit cell, $\alpha$ and $\beta$ are the
Cartesian coordinates $x,y,z$, $f$ is the Fermi distribution and finally $\underline{\underline{v}}_{\alpha}(\mathbf{k})$ denote the Fermi velocities obtained from the LDA Hamiltonian $\underline{\underline{\varepsilon}}^{\rm LDA}(\mathbf{k})$.
The spectral function $\underline{\underline{A}}$ reads\\
\begin{equation}\label{specfunc}
\underline{\underline{A}}(\mathbf{k},\omega)=-\frac{1}{\pi}{\rm Im}\left[(\omega+\mu)\underline{\underline{\mathbbm{1}}}-\underline{\underline{\varepsilon}}^{\text{LDA}}(\mathbf{k})-\underline{\underline{\Sigma}}(\omega)\right]^{-1} \;\;{\rm .}
\end{equation}\\
Here, the double underlines expresses the matrix character in spin and orbital
space.  Hence, the correlation effects enter Eq.~\eqref{finalopt}
only via the DMFT self energy $\underline{\underline{\Sigma}}(\omega)$ in
$\underline{\underline{A}}(\mathbf{k})$. Let us also note that due to the trace in
Eq.~\eqref{finalopt} orbital off diagonal contributions are fully taken
into account. Finally we remark that in expression~\eqref{finalopt} vertex corrections are
neglected. For the single band case, this
becomes exact in the limit of infinite dimensions, just like DMFT does. In a multi-band
system with similar orbitals, such as V$_2$O$_3$ it is a good approximation (Tomczak, J.~M., and Biermann, S., {\it Physical Review B}  \textbf{80}, 085127 (2009)).}\\

\noindent\Large{Calculation of $\underline{\underline{\Sigma}}$ and $\underline{\underline{A}}$}\\

\noindent\small{Our LDA+DMFT calculations were carried out on the crystal structure 
of V$_{0.989}$Cr$_{0.011}$)$_2$O$_3$ in the $\alpha$ and $\beta$ phases
( Robinson, W.~R.,
{\it Acta Crystallographica Section B}, \textbf{31}, 1153 (1975)
). The LDA 
calculations were performed in the localized basis of muffin tin
orbital (MTO) based LMTO and NMTO ( Andersen, O. K. and Jepsen, O., { \it Phys. Rev. Lett.}, {\textbf 53}, 2571 (1984); Andersen, O. K. {\it Lecture Notes in Physics Springer, New York, 2000)}).
The full LDA Hamiltonian was downfolded on to the 
V t2g only low-energy Hamiltonian, defined in the effective Wannier function representation, 
using the NMTO-downfolding technique. This  yields a 12 $\times$ 12
Hamiltonian for each k-point of the Brillouin Zone
as there are four vanadium
atoms in the primitive unit cell. For a detailed discussion of the 
downfolding procedure and the setup of the LDA+DMFT calculation (Saha-Dasgupta, T., et al,
{\it arXiv.org,} \textbf{0}, 0907.2841 (2009); 
Ref.26, main paper). For further details about LDA+DMFT see Held, K., {\it Advances in Physics}, \textbf{56}, 829  (2007).
To obtain the self energy matrix $\underline{\underline{\Sigma}}(\omega)$ we first perform an analytical
continuation of the local Green function
$\underline{\underline{G}}^{\text{loc.}}(\omega)$ using the maximum entropy
code by Jarrell and Gubernatis (Jarrell, M. and Gubernatis, J.~E., 
{\it Physics Reports}, \textbf{269}(3), 133 (1996)
) and some improvements described by Gunnarson, Haverkort, and Sangiovanni (Gunnarsson, O.,  Haverkort, M.W.,  and Sangiovanni, G.,{\it Phys. Rev. B} {\bf 81}, 155107 (2010)). In a second step we solve
a multidimensional rootfinding problem\\
\begin{equation}
\underline{\underline{G}}^{\text{loc.}}(\omega)=\frac{1}{V_{\text{BZ}}}\int_{\text{BZ}}d^3k\frac{1}{(\omega+\mu)\underline{\underline{\mathbbm{1}}}-\underline{\underline{\varepsilon}}^{\text{LDA}}(\mathbf{k})-\underline{\underline{\Sigma}}(\omega)} 
\end{equation}

In Supplementary Fig.S2 we report the orbital-resolved LDA+DMFT self energies. In the plots we set the Fermi energy to
$\varepsilon_{\text{F}}=0$ and plot the 
$e_g^\pi$ self energy in green and the $a_{1g}$ self energy in blue. We
summarize the quantities for the $\alpha$- and the $\beta$-phase on
the left hand and right hand side of the panels respectively. Overall
our results agree with the results of the previous LDA+DMFT 
analysis (Ref. 26, main paper), although we
performed the calculations at slightly lower $U=4.0$ eV values (see Ref.26,  $U=4.2$ eV) in order to get the best agreement with
experiment. Indeed we found an extreme sensitivity of the optical gap
depending on the value of the Hubbard $U$. For a detailed analysis of
the $U$ dependence of the LDA+DMFT calculations see
 Toschi, A.,  et al,
 {\it Journal of Physics: Conference Series}, \textbf{200}, 012208 (4pp)
  (2010).}

\renewcommand{\figurename}{{\bf Supplementary Fig.S}}
\begin{figure}[h]   
\begin{center}    
\includegraphics[width=12cm]{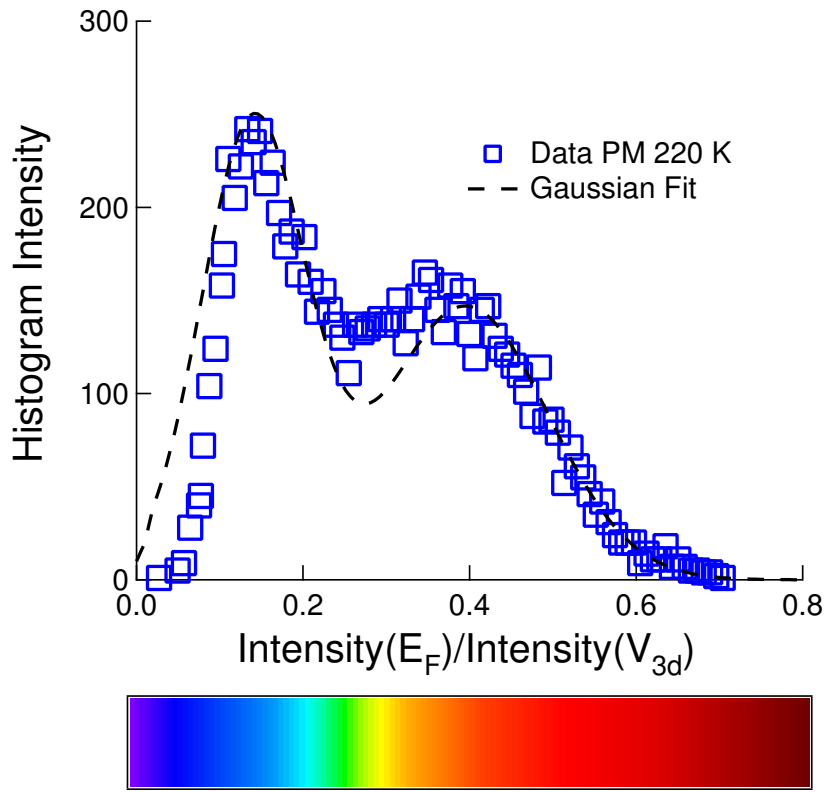} 
\caption{ {\bf Intensity distribution of PM photemission map.} 
Histogram of the intensity distribution of the PM photoemission map (ratio between the photoemission intensities at E$_{F}$ and on the V3d band) (open symbols) as extracted from the map at 220 K in Fig.2a in the paper. A bimodal distribution can be observed. This distribution was fitted with two gaussian components (dashed line) providing f=0.45$\pm$0.05. The color scale on the horizontal axis is the same used in the photoemission maps in Fig. 2 of the paper.
} 
\label{Fig.1_Suppl}
\end{center}
\end{figure}

\begin{figure}
  \begin{center}
  \includegraphics[width=14cm]{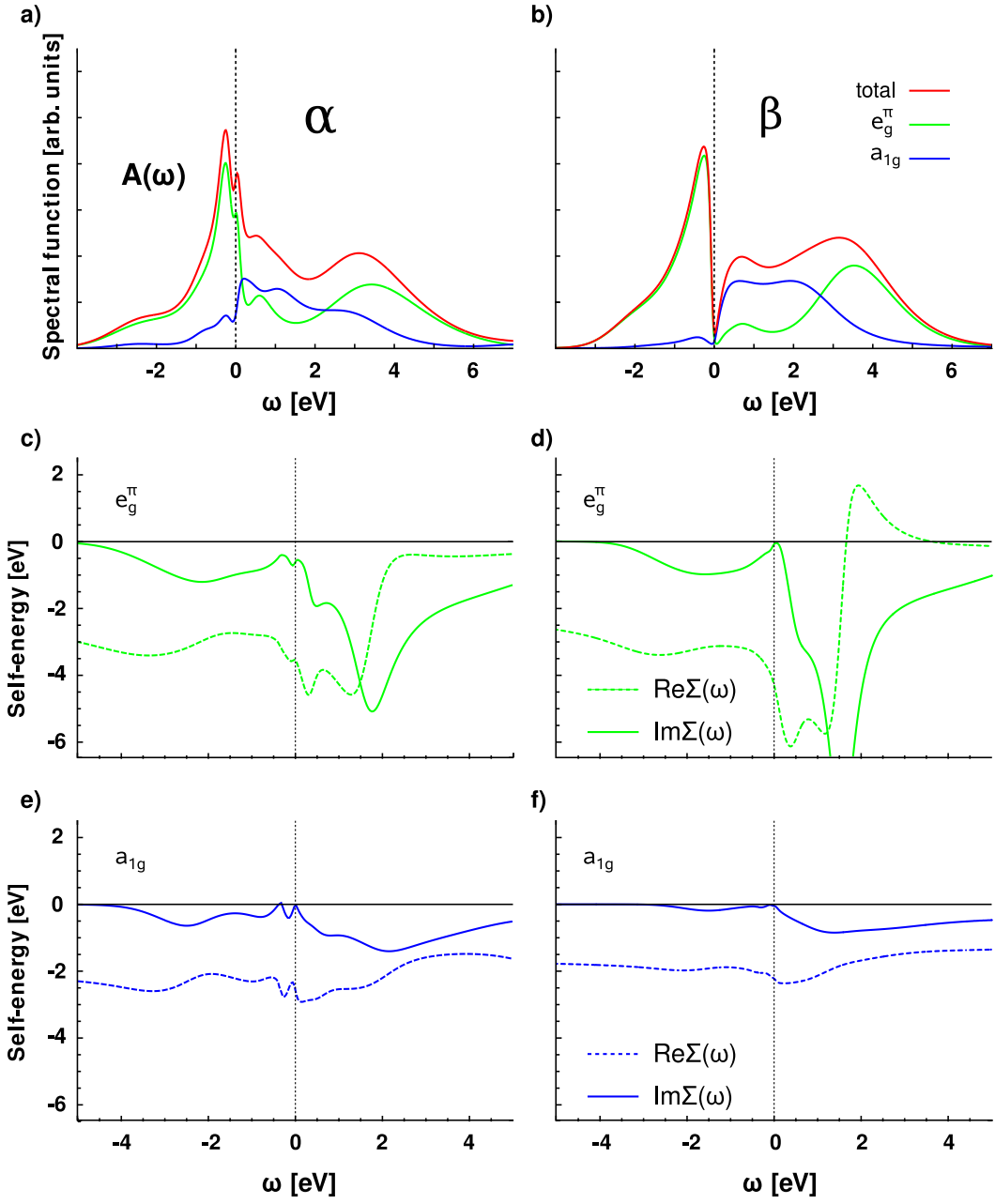}
  \end{center}
    \caption{ {\bf LDA+DMFT self energy.} 
    LDA+DMFT self energy for (V$_{0.989}$Cr$_{0.011}$)$_2$O$_3$ in
      the $\alpha$ phase (left hand side) and the $\beta$ phase (right
      hand side) for $U=4.0$ eV and $T=390K$. In the top and bottom panels  we show the LDA+DMFT self energy
      for the $e_g^\pi$ and $a_{1g}$ states respectively.}
    \label{V2O3_awsfspec}
\end{figure}

\end{document}